\documentclass[twocolumn,conference,letterpaper]{IEEEtran}
\usepackage[left=0.75in,right=0.75in,top=1in,bottom=0.75in]{geometry}

\usepackage{cite}
\usepackage{amssymb,amsmath,latexsym,amsfonts,amsthm,mathtools}

\usepackage{graphicx}
\usepackage{float,subcaption}
\usepackage{textcomp}
\usepackage{xcolor}
\definecolor{darkpastelpurple}{rgb}{0.59, 0.44, 0.84}
\usepackage{hyperref}
\hypersetup{colorlinks, breaklinks, citecolor=blue, linkcolor=blue, urlcolor=blue}
\usepackage[nameinlink]{cleveref}
\Crefformat{figure}{#2Fig.~#1#3}
\Crefmultiformat{figure}{Figs.~#2#1#3}{ and~#2#1#3}{, #2#1#3}{ and~#2#1#3}

\usepackage{tikz}
\usetikzlibrary{automata, shapes, arrows, calc, arrows.meta, fit, positioning}

\theoremstyle{plain}
\newtheorem{theorem}{Theorem}
\newtheorem{lemma}{Lemma}

\newtheorem*{problem*}{Problem}
\theoremstyle{remark}
\newtheorem{remark}{Remark}
\newtheorem{assumption}{Assumption}

\theoremstyle{definition}

\newcommand{\sign}{\mathrm{sign}}
\usepackage{mathrsfs}

\usepackage{newtxmath}
\usepackage{booktabs}
\usepackage{duckuments}

\newcommand{\vect}[1]{\mathbf{#1}}
\newcommand{\augstate}{\tilde{\mathbf{x}}}

\usepackage{accents}

\IEEEoverridecommandlockouts

\begin{document}

\title{Robust Near-Optimal Nonlinear Target Enclosing Guidance} 

\author{Abhinav Sinha,~\IEEEmembership{Senior Member,~IEEE}, and Rohit V. Nanavati, \IEEEmembership{Member,~IEEE}
 \thanks{A. Sinha is with the Guidance, Autonomy, Learning, and Control for Intelligent Systems (GALACxIS) Lab, Department of Aerospace Engineering and Engineering Mechanics, University of Cincinnati, OH 45221, USA. (e-mail: abhinav.sinha@uc.edu).
 R. V. Nanavati is with the Autonomous Intelligent Robotics (AIR) Lab, Department of Aerospace Engineering, Indian Institute of Technology Bombay, Mumbai 400076, India. (e-mail: r.nanavati@aero.iitb.ac.in).}}
\maketitle
	
\begin{abstract}
This paper proposes a nonlinear optimal guidance law that enables a pursuer to enclose a target within arbitrary geometric patterns, which extends beyond conventional circular encirclement. The design operates using only relative state measurements and formulates a target enclosing guidance law in which the vehicle’s lateral acceleration serves as the steering control, making it well-suited for aerial vehicles with turning constraints. Our approach generalizes and extends existing guidance strategies that are limited to target encirclement and provides a degree of optimality. At the same time, the exact information of  the target's maneuver is unnecessary during the design. The guidance law is developed within the framework of a state-dependent Riccati equation (SDRE), thereby providing a systematic way to handle nonlinear dynamics through a pseudo-linear representation to design locally optimal feedback guidance commands through state-dependent weighting matrices. While SDRE ensures near-optimal performance in the absence of strong disturbances, we further augment the design to incorporate an integral sliding mode manifold to compensate when disturbances push the system away from the nominal trajectory, and demonstrate that the design provides flexibility in a sense that the (possibly time-varying) stand-off curvature could also be treated as unknown. Simulations demonstrate the efficacy of the proposed approach.
\end{abstract}


\section{Introduction}\label{sec:introduction}
    Driven by increasing autonomy requirements, unmanned aerial vehicles (UAVs) have become integral to a variety of mission-critical operations such as area coverage, surveillance, environmental monitoring, aerial defense, reconnaissance, and search and rescue, e.g., \cite{nanavati2024distributed,kumar2024robust,SINGH2022592,Sinha2022}. A common control objective in these applications involves regulating the motion of a UAV designated as the pursuer to maintain a prescribed geometric configuration or trajectory with respect to a reference entity, termed the target. The target may correspond to another vehicle, a stationary beacon, or a dynamically evolving point of interest. This objective, referred to as target enclosing, entails enforcing spatial constraints on the relative position of the pursuer with respect to the target, and is central to both civilian and military guidance and coordination strategies.
    
    One of the earliest investigations into target enclosing using teams of mobile robots was presented in \cite{yamaguchi}, where the agents were modeled under the assumption of holonomic dynamics. Subsequent works extended this framework, leveraging the relative simplicity of control synthesis afforded by holonomic vehicle models. However, in practical scenarios, the dynamics of mobile platforms are more accurately represented by non-holonomic constraints, which significantly complicate both analysis and control design. Target enclosing under non-holonomic vehicle dynamics has thus been examined in a more realistic setting in works such as \cite{ZHENG2013401,CECCARELLI20083025}, often within the broader context of distributed formation control. Within this domain, a task of interest is maintaining circular motion around a target (circumnavigation). This specific formation behavior has attracted substantial research attention due to its relevance in surveillance, tracking, and protection missions, as evidenced by a wide body of literature (e.g., \cite{9924233,9462350,MATVEEV2011177,6621827,6060869}).
    
    One approach is to leverage vector fields defined around the desired trajectory or geometric pattern to regulate the vehicle’s heading and generate convergence behavior. In \cite{doi:10.2514/1.30507}, the authors employed a Lyapunov-based vector field to coordinate the circumnavigation of a stationary target at multiple altitudes using a fleet of UAVs. The work in \cite{doi:10.2514/1.37212} extended this methodology to account for external disturbances, proposing a coordinated standoff tracking strategy in the presence of wind. In contrast to purely reactive vector field approaches, the work in \cite{doi:10.2514/1.56254} presented a nonlinear model predictive control framework for dual-vehicle coordinated standoff tracking, optimizing performance over a finite prediction horizon. An alternative formulation in \cite{6621827} utilized spherical pendulum-inspired dynamics, wherein backstepping and Lyapunov-based techniques were applied to regulate position and velocity tracking errors for robust guidance law synthesis. Beyond vector field and predictive control approaches, several recent strategies have explored guidance law design under partial or minimal state information, often motivated by scenarios with limited sensing or communication capabilities. These methods typically employ either relative range and or bearing measurements (see, for example, \cite{6060869,MATVEEV2011177,6705614,8798252,9924233}).
    
    Despite such solutions, several key challenges remain unaddressed. This work advances the state of the art in target-enclosing guidance by addressing limitations of existing methods, which we summarize below:
    \begin{itemize}
        \item Many of the aforementioned methods assume slow or negligible target motion, which may become limited under adversarial scenarios if the target moves aggressively. The proposed framework explicitly accommodates a general maneuvering target without requiring knowledge of its guidance law or future maneuvers.
        \item While previous studies rely on curvature information of the enclosing geometry, our formulation has the flexibility to eliminate this requirement, thereby broadening applicability to arbitrary geometric patterns provided the prescribed range-parameterized curve is at least $\mathbb{C}^2$.
        \item We introduce a robust nonlinear optimal guidance design for this problem class by integrating an SDRE framework with a supertwisting sliding mode controller defined over an integral sliding manifold. This hybrid design leverages the optimality of SDRE while inheriting robustness and finite-time convergence properties from sliding mode control, where the reaching phase is absent.
        % \item Comparative simulations against a representative state-of-the-art method demonstrate that our approach achieves target enclosing with significantly reduced control energy.
    \end{itemize}

    \section{Problem Formulation}
\begin{figure}[ht!]
        \centering
        \includegraphics[width=0.75\linewidth]{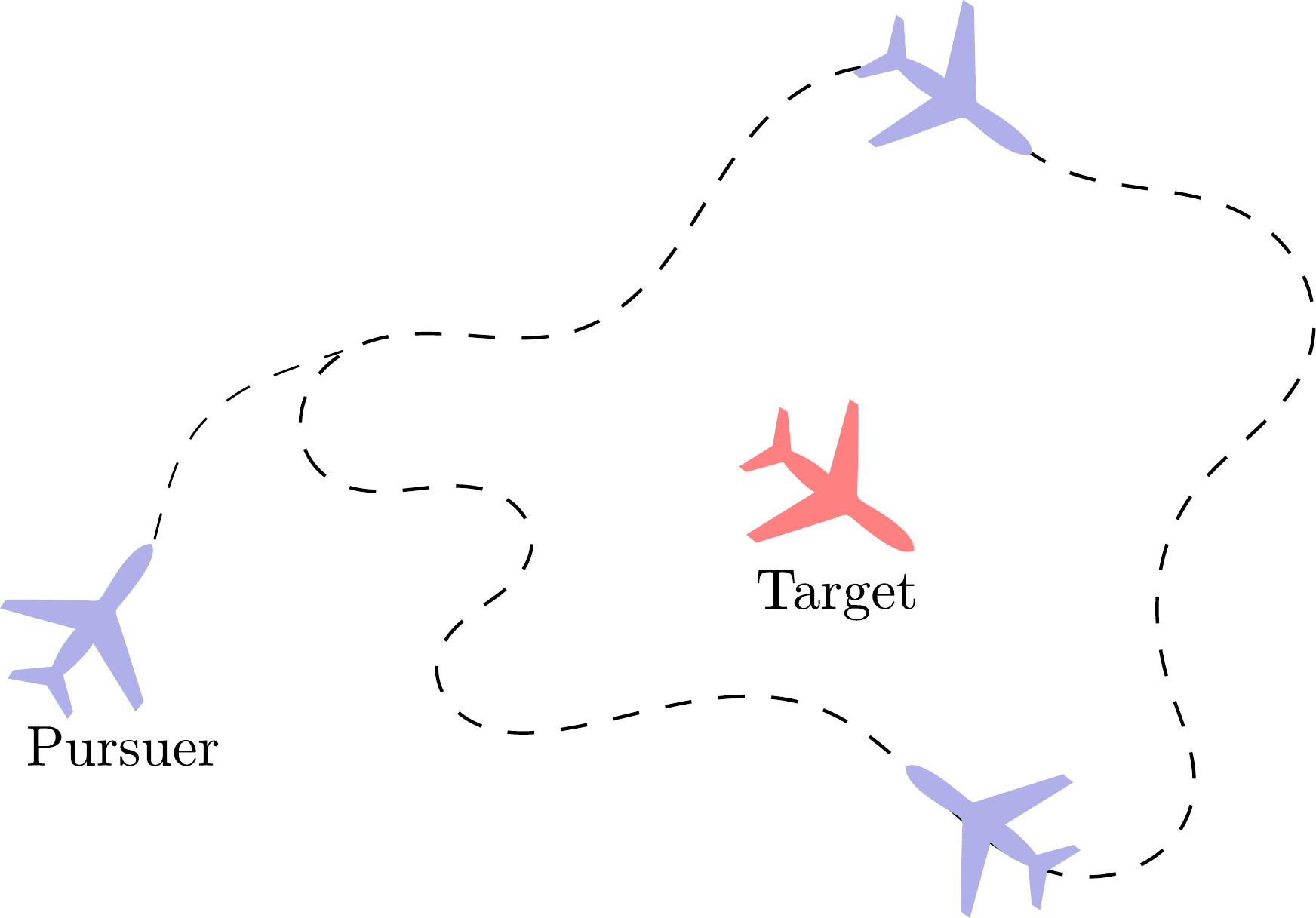}
        \caption{Target enclosing within a circle.}
			\label{fig:encircle}
    \end{figure}

\begin{figure}[h!]
		\centering
		\includegraphics[width=\linewidth]{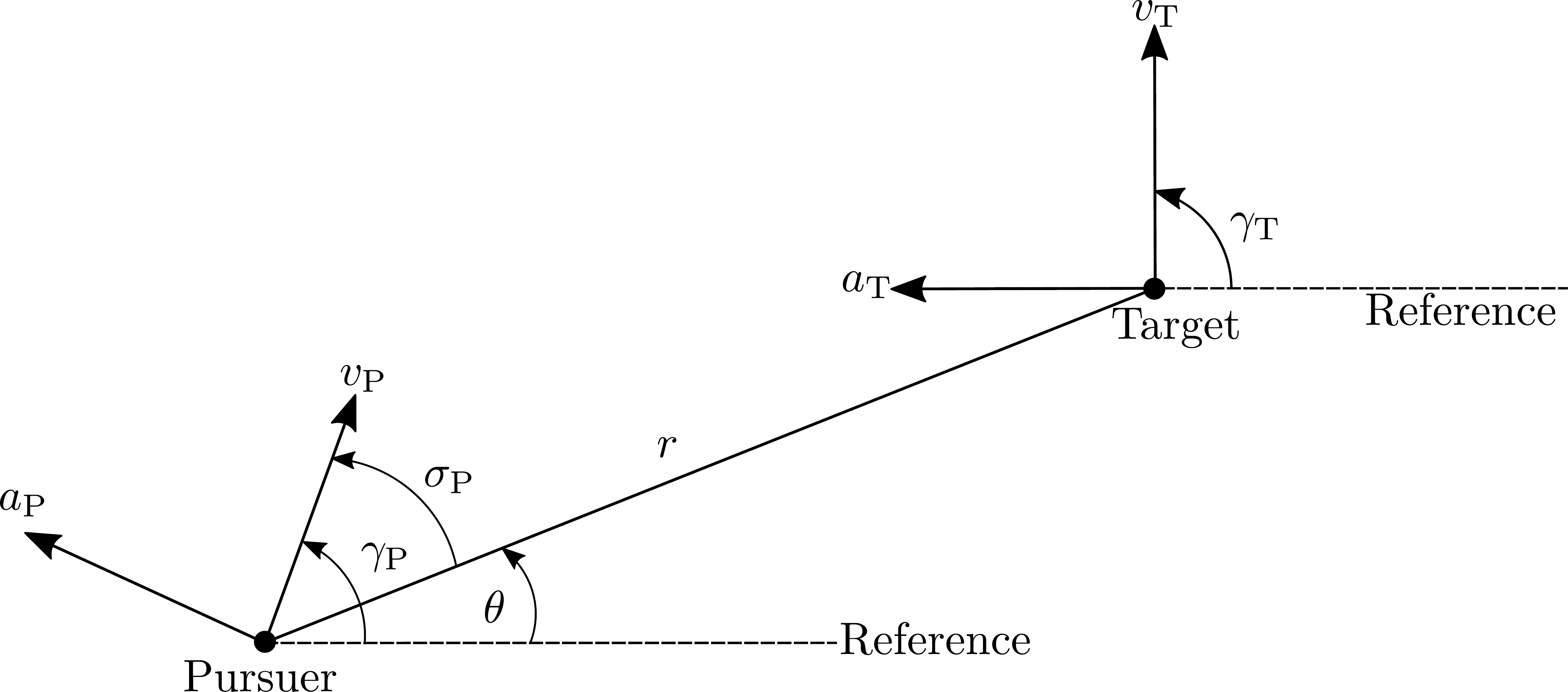}
			\caption{Planar engagement geometry.}
			\label{fig:enggeo}
	\end{figure}

    Consider a scenario in which a pursuing vehicle seeks to enclose a target within a prescribed geometric pattern, as illustrated in \Cref{fig:encircle}.
    The engagement geometry associated with this setup is depicted in \Cref{fig:enggeo}. The pursuer’s translational velocity and flight path angle are denoted by $v_\mathrm{P}$ and $\gamma_\mathrm{P}$, respectively. The pursuer is steered via its lateral acceleration $a_\mathrm{P}$ applied perpendicular to its velocity vector, consistent with a planar, nonholonomic kinematic model. Let $r$ denote the relative range (line-of-sight separation) between the pursuer and the target, and let $\theta$ be the line-of-sight (LOS) angle defined with respect to an inertial reference frame. The look angle $\sigma_\mathrm{P}=\gamma_\mathrm{P}-\theta$ represents the relative orientation between the pursuer’s heading and the LOS vector. Without loss of generality, we assume that the target is also mobile and governed by similar kinematic constraints as the pursuer. Specifically, the target has a speed $v_\mathrm{T}$, heading angle $\gamma_\mathrm{T}$, and lateral acceleration $a_\mathrm{T}$, which is unknown to the pursuer but remains bounded, i.e. $|a_\mathrm{T}|\leq a_\mathrm{T}^{\max}$.
    \begin{assumption}
        Both the pursuer and the target are modeled as point-mass vehicles without the capability to accelerate radially, and $v_\mathrm{P}>v_\mathrm{T}$.
    \end{assumption}
    The relative motion between the pursuer and the target, expressed in polar coordinates with the target serving as the reference point (\Cref{fig:enggeo}), is governed by
    	\begin{subequations}\label{eq:engdyn}
    		\begin{align}
    			v_r = \dot{r} =& v_\mathrm{T} \cos\left(\gamma_\mathrm{T}-\theta\right) - v_\mathrm{P} \cos\sigma_\mathrm{P}, \label{eq:rdot} \\
    			v_\theta  =r\dot{\theta} =& v_\mathrm{T} \sin\left(\gamma_\mathrm{T}-\theta\right) - v_\mathrm{P} \sin\sigma_\mathrm{P}, \label{eq:thetadot}\\
    			\dot{\gamma}_i =& \dfrac{a_i}{v_i};\;\left\vert a_i\right\vert \leq a_i^\mathrm{max},\,\forall\,i\in\left\{\mathrm{P},\mathrm{T}\right\},\label{eq:gammadot}
    		\end{align}
    	\end{subequations}
        \begin{remark}
            By formulating the steering control in terms of lateral acceleration, the vehicles are modeled as turn-constrained systems, i.e., they possess a finite minimum turning radius and cannot instantaneously change their direction of motion.
        \end{remark}
    \begin{problem*}
        Let the desired enclosing geometry be defined by a smooth, time-varying, closed curve $\mathcal{C}(t)\subset\mathbb{R}^2$ parameterized by a reference point $r_d(t)$. The goal is to design a robust optimal nonlinear feedback control law (the guidance strategy) for the pursuer's steering input (lateral acceleration) such that $\lim_{t\to\infty}|r(t)-r_d(t)|\to 0$.
    \end{problem*}

\section{Main Results}
    To facilitate the control design, we exercise control over the relative range. Note that the guidance problem can be cast as an output regulation problem in terms of the range error.
    \begin{lemma}\label{lem:reldeg}
        Let us define the range error variable as $\rho = r - r_d$, whose dynamics possess relative degree two with respect to the pursuer’s lateral acceleration $a_\mathrm{P}$.
    \end{lemma}
    \begin{proof}
    Differentiating $\rho$ with respect to time yields
    \begin{align}
    \dot{\rho} &= \dot{r} - \dot{r}_d = v_\mathrm{T} \cos(\gamma_\mathrm{T} - \theta) - v_\mathrm{P} \cos\sigma_\mathrm{P}  - \dot{r}_d, \label{eq:epsdot}
    \end{align}
    where \eqref{eq:rdot} has been used to substitute for $\dot{r}$. Differentiating once more, one may obtain
    	\begin{align}
    			\ddot{\rho} =&~ -v_\mathrm{T} \sin\left(\gamma_\mathrm{T}-\theta\right)\,\left(\dfrac{a_\mathrm{T}}{v_\mathrm{T}}-\dot{\theta}\right) + v_\mathrm{P} \sin\sigma_\mathrm{P} \left(\dfrac{a_\mathrm{P}}{v_\mathrm{P}}-\dot{\theta}\right) - \ddot{r}_d \nonumber\\
    			=&~ \left( v_\mathrm{T} \sin\left(\gamma_\mathrm{T}-\theta\right) - v_\mathrm{P} \sin\sigma_\mathrm{P}\right)\dot{\theta} 
                + a_\mathrm{P}\sin\sigma_\mathrm{P} \nonumber\\ &- a_\mathrm{T}\sin\left(\gamma_\mathrm{T}-\theta\right)- \ddot{r}_d,\label{eq:eddotstep2}
    		\end{align}
    		which can be expressed, using \eqref{eq:thetadot}, as
    		\begin{align}
    			\ddot{\rho} =&~ r\dot{\theta}^2 + a_\mathrm{P}\sin\sigma_\mathrm{P} - a_\mathrm{T}\sin\left(\gamma_\mathrm{T}-\theta\right)- \ddot{r}_d,\label{eq:eddot}
    		\end{align}
         showing $a_\mathrm{P}$ influences $\rho$ with a relative degree of two.
    \end{proof}
    It is evident from \Cref{lem:reldeg} that the desired standoff distance must satisfy a regularity assumption ($r_d(t)\in\mathbb{C}^2$) in the sense that an unbounded or highly irregular reference trajectory does not correspond to a meaningful enclosing objective in practical scenarios. Therefore, we impose the following boundedness conditions on the desired relative range, that is, $0<r_d^{\min}\leq r_d\leq r_d^{\max}, ~|\dot{r}_d|\leq\dot{r}_d^{\max}<\infty,~|\ddot{r}_d|\leq\ddot{r}_d^{\max}<\infty$. 
    
    Using the results in Lemma \ref{lem:reldeg}, we can construct a state space model representing the dynamics of the error $\rho$ as
    \begin{align}\label{eq:dynamics}
        \dot{\vect{x}} = \vect{f}(\vect{x}) + \vect{B}(\vect{x})a_\mathrm{P} + \vect{h}(\vect{x})
    \end{align}
    with state vector $\vect{x} = \begin{bmatrix}
        x_1 & x_2
    \end{bmatrix}^\top = \begin{bmatrix}
        \rho & \dot{\rho}
    \end{bmatrix}^\top$, and drift dynamics $\vect{f}(\vect{x}) = \begin{bmatrix}
            x_2 &
            x_1\dot{\theta}^2 + \vect{g}(\vect{x})
        \end{bmatrix}^\top$. The state-dependent function $\vect{g}(\vect{x})={r}_d\dot{\theta}^2 - \ddot{r}_d,$ captures the reference curvature. The control input and the disturbance matrices take the form
    \begin{gather}
        \vect{B}(\vect{x}) = \begin{bmatrix}
            0 \\ \sin\sigma_\mathrm{P}
        \end{bmatrix},~~\vect{h}(\vect{x}) = \begin{bmatrix}
                0\\
                - a_\mathrm{T}\sin(\gamma_\mathrm{T}-\theta)
            \end{bmatrix}.
    \end{gather}
    so that $a_\mathrm{P}$ enters linearly in the second channel, consistent with the relative degree established in Lemma~\ref{lem:reldeg}.  
    \begin{remark}\label{rmk:gxhx}
       If the function $\mathbf{g}(\mathbf{x})$ is contained in $\vect{h}(\vect{x})$, then it essentially means that the curvature information is not explicitly available in the system representation. For simplicity, here we assume that such information is known. However, note that the proposed design can handle the cases where $\mathbf{g}(\mathbf{x})$ is a part of $\vect{h}(\vect{x})$, thus broadening applicability to arbitrary geometric patterns provided the prescribed range-parameterized curve is at least $\mathbb{C}^2$.
    \end{remark}

    We seek to factor $\vect{f}(\vect{x})$ as $\vect{A}(\vect{x})\vect{x}$ in order to cast the nonlinear dynamics above into the SDRE pseudo-linear structure, which is a prerequisite for applying Riccati-based optimal control design. This necessitates that the drift vector field satisfy $\vect{f}(\vect 0) = \vect{0}$. However, in our case $\vect{f}(\vect{0})\neq\vect 0$ since $\vect g(\vect0)\neq 0$.  To remedy this issue, we reconstruct the state-space equations by isolating the non-vanishing offset term into the disturbance vector. This yields a modified representation in which the drift term is linear in the state, while the residual nonlinearities and constant components are absorbed into $\vect{h}(\vect{x})$ \cite{Kumar2020,CIMEN20083761}. The resulting model admits the standard SDRE structure. To this end, we introduce an auxiliary variable $\eta$ leading to 
    \begin{align}
        \dot{\augstate} &= \mathcal{A}(\augstate)\augstate + \mathcal{B}(\augstate)a_\mathrm{P} + \mathcal{H}(\augstate),\label{eq:augdynamics}\\
        \vect{y} &= \mathcal{C}\augstate\label{eq:augouput}
    \end{align}
    where $\augstate = \begin{bmatrix}
        \vect{x}^\top & \eta
    \end{bmatrix}^\top$, $\vect{y}$ is the output with $\mathcal{C} = \begin{bmatrix}
            \mathbf{I}_{2\times2} & \mathbf{0}
        \end{bmatrix}$. The matrices $\mathcal{A}(\augstate)$ and $\mathcal{B}(\augstate)$ are defined as
    \begin{subequations}
    \begin{gather}\label{eq:system_and_input}
        \mathcal{A}(\augstate) = \begin{bmatrix}
            \vect{A}(\vect{x}) & \vect{G}(\vect{x})\\0& -\Lambda
        \end{bmatrix},~~\mathcal{B}(\augstate) = \begin{bmatrix}
            \vect{B}(\vect{x})\\
            0
        \end{bmatrix},\\
        \vect{A}(\vect{x}) = \begin{bmatrix}
            0 & 1\\
            \dot{\theta}^2 & 0
        \end{bmatrix},~~
        \vect{G}(\vect{x}) = \begin{bmatrix}
            0 \\
            \frac{\vect{g}(\vect{x})}{\eta}
        \end{bmatrix},~~
        \mathcal{H}(\augstate) = \begin{bmatrix}
            \vect{h}(\vect{x}) \\
            0
        \end{bmatrix},
    \end{gather}
    \end{subequations}
    where $0<\Lambda<<1$ and the auxiliary variable satisfies $\eta(0)\ne0$.     %\tr{Is $\Lambda$ very small than 1? We also need to streamline the notations. If there are vector zeroes, then we need to include them in boldface.}

    In order to drive the output, \eqref{eq:augouput}, to zero subject to \eqref{eq:augdynamics}, we design the guidance strategy based on SDRE and super-twisting sliding mode control with an integral sliding manifold, as $a_\mathrm{P} = a_{\mathrm{P},n} + a_{\mathrm{P},d}$. The term $a_{\mathrm{P},n}$ denotes a near-optimal nonlinear feedback control input designed to drive the nominal system \eqref{eq:dynamics} with $\vect{h}(\vect{x})=0$ to the origin while minimizing the cost function $J$ defined as
    \begin{align}
        J &= \int_{t_0}^{\infty}\left[\vect{y}^\top\vect{Q}\vect{y} + Ra_\mathrm{P}^2\right]d\tau = \int_{t_0}^{\infty}\left[\augstate^\top C^\top\vect{Q}C\augstate + Ra_\mathrm{P}^2\right]d\tau.\label{eq:costfunction}
    \end{align}
    In \eqref{eq:costfunction}, the weighting parameters satisfy $\vect{Q}\in\mathbb{R}^{2\times2}\succeq 0$ (positive semidefinite) and $R\succ 0$ (positive definite) for all $\augstate$. The term $a_{\mathrm{P},d}$ denotes the disturbance rejection input to handle the unknown target maneuver modeled in $\vect{h}(\vect{x})$. Such partitioning has also appeared, for example, in \cite {Kumar2020}.

    \begin{lemma}\label{lem:controllability}%\cite{Kumar2020}
        The pairs $\{\mathcal{A},\mathcal{B},\mathcal{C}\}$ and $\{\mathcal{A},\mathcal{Q} = \sqrt{C^\top QC}\}$ are point-wise output stabilizable and state detectable parameterization, respectively, of the system \eqref{eq:augdynamics}-\eqref{eq:augouput}, in the linear sense, $~\forall~\augstate$ with $\sigma_\mathrm{P}\notin \{0,\pm\pi\}$.
    \end{lemma}
    \begin{proof}
        Define the output controllability matrix for the system \eqref{eq:augdynamics}-\eqref{eq:augouput}, in the linear sense, as $\mathcal{O}_c = \left[\mathcal{C}\mathcal{B}~~\mathcal{C}\mathcal{AB}~~\mathcal{C}\mathcal{A}^2\mathcal{B}\right]$\,. On substituting \eqref{eq:system_and_input} in $\mathcal{O}_c$, we get $$\mathcal{O}_c = \begin{bmatrix}
            0 & \sin\sigma_\mathrm{P} & 0\\
            \sin\sigma_\mathrm{P} & 0 & \dot{\theta}\sin\sigma_\mathrm{P}
        \end{bmatrix}\,.$$
        It can be easily be noted that the $\text{\rm rank}(\mathcal{O}_c)=2~~\forall~\sigma_\mathrm{P}\notin\{0,\pm\pi\}$ implying output stabilizability for all $\augstate$ for which $\sigma_\mathrm{P}\notin\{0,\pm\pi\}$. Similarly, it can be shown that the system \eqref{eq:augdynamics}-\eqref{eq:augouput} is state detectable as the state detectability matrix $\mathcal{O}_o = \left[\mathcal{Q}^\top~~\mathcal{A}^\top\mathcal{Q}^\top~~(\mathcal{A}^2)^\top\mathcal{Q}^\top\right]^\top$  has full column rank. 
    \end{proof}
    \begin{remark}
        It can be inferred from \eqref{eq:dynamics} that $\sigma_\mathrm{P}\in\{0,\pm\pi\}$ may be an input singularity as $\vect{B}=0$ for $\sigma_\mathrm{P}\in\{0,\pm\pi\}$. However, such singular points are isolated and belong to a set of measure zero. During implementation, this does not pose an issue.
    \end{remark}
    \begin{remark}
        The controller input, $a_{\mathrm{P},n}$, that minimizes \eqref{eq:costfunction} subject to the system \eqref{eq:augdynamics}-\eqref{eq:augouput}, can be obtained as 
    \begin{align}\label{eq:nominal_controller}
        a_{\mathrm{P},n} = -R^{-1}\mathcal{B}^\top \mathcal{P}(\augstate)\augstate\,,
    \end{align}
    where the matrix $\mathcal{P}(\augstate)\succeq 0\in\mathbb{R}^{3\times3}$ is computed by solving the algebraic Riccati equation \cite{CIMEN20083761,Cimen2012,Kumar2020} given by 
    \begin{align}\label{eq:ARE}
            \mathcal{A}^\top\mathcal{P} + \mathcal{P}\mathcal{A} - \mathcal{P}\mathcal{B}R^{-1}\mathcal{B}^\top\mathcal{P} + \mathcal{C}^\top\vect{Q}\mathcal{C} = 0\,.
    \end{align}
    Note that as \eqref{eq:ARE} is an algebraic equation consisting of state-dependent matrices $\mathcal{A}$ and $\mathcal{B}$, the nominal lateral acceleration is computed by recursively solving \eqref{eq:ARE} along $\augstate(t)$ in the feedback sense. 
    \end{remark}
    We now present the design of the disturbance rejection input in the next theorem.
    \begin{theorem}\label{thm:guidance}
        Consider the system \eqref{eq:augdynamics}–\eqref{eq:augouput} and define the sliding manifold
        \begin{align}
            \mathscr{S} = \vect{L}\int_{0}^{t}\left[\dot{\vect{y}} - \mathcal{C}\left\{\mathcal{A}(\augstate) -\mathcal{M}(\augstate)\mathcal{P}(\augstate)\right\}\augstate\right]d\tau\label{eq:switchingsurface}
        \end{align}
        where $\mathcal{M} = \mathcal{B}(\augstate) R^{-1}\mathcal{B}^\top(\augstate)$ and $\vect{L}$ is a the design matrix. Let the pursuer's lateral acceleration input be $a_\mathrm{P} = a_{\mathrm{P},n} + a_{\mathrm{P},d}$,
        % \begin{align}\label{eq:guidance_command}
        %     a_\mathrm{P} = a_{P,n} + a_{P,d}
        % \end{align}
        with the nominal component $a_{\mathrm{P},n}$ computed using \eqref{eq:nominal_controller} and 
        \begin{align}
            a_{\mathrm{P},d} = \left[\vect{L}\mathcal{C}\mathcal{B}\right]^{-1}\left(-\alpha_1|\mathscr{S}|^{\beta}\circ\sign(\mathscr{S}) + \vect{w}\right)
        \end{align}
        with $\alpha_1,\beta\in\mathbb{R}_+$, the auxiliary dynamics satisfies $\dot{\vect{w}} = -\alpha_2\sign(\mathscr{S})$ for some $\alpha_2\in\mathbb{R}_+$, and $\circ$ denotes the element-wise power times sign on scalar arguments. If the design matrix $\vect{L}$ is chosen such that $\vect{L}\mathcal{C}\mathcal{B}$ is invertible, then the closed-loop output \eqref{eq:augouput}, subject to \eqref{eq:augdynamics}, is locally exponentially stable. 
    \end{theorem}
    \begin{proof}
    The proof is carried out in two steps. In the first step, we show the stability of the sliding mode dynamics, followed by the stability analysis during the sliding phase.

    Define a Lyapunov function candidate \begin{align}
        V_1(\mathscr{S},\mathbf{w}) =& \dfrac{1}{2}\begin{bmatrix}
        \sqrt{|\mathscr{S}|+\epsilon}\circ\sign(\mathscr{S})&\mathbf{w}
    \end{bmatrix}^\top\nonumber\\
    &\times\mathscr{P}\begin{bmatrix}
        \sqrt{|\mathscr{S}|+\epsilon}\circ\sign(\mathscr{S})&\mathbf{w}
    \end{bmatrix},
    \end{align}
    where $\mathscr{P}\succ 0$ is constant and $\mathscr{P} = \mathscr{P}^\top$, and $\epsilon>0$ is a small regularization constant to avoid singularity in $\mathscr{S}_i$. Let $\varrho = \begin{bmatrix}
        \sqrt{|\mathscr{S}|+\epsilon}\circ\sign(\mathscr{S})&\mathbf{w}
    \end{bmatrix}^\top$. Then, one may express $\dot{V}_1$ as $\dot{V}_1 = \varrho^\top \mathscr{P}\dot{\varrho} + \dot{\varrho}^\top\mathscr{P}\varrho$, where $\dot{\varrho}$ is needed for further simplification. To this end, we differentiate $\varrho$ to obtain $\dot{\varrho} = \left(|\mathscr{S}|+\epsilon\right)^{-\frac{1}{2}}\circ \left(\boldsymbol{\Psi} \varrho + \vect{L}\mathcal{C}\mathcal{H}(\augstate)\right)$,
	% \begin{align}
	% \dot{\varrho} =& |\mathscr{S}|^{-\frac{1}{2}}\circ \left(\boldsymbol{\Psi} \varrho + \vect{L}\mathcal{C}\mathcal{H}(\augstate)\right),%\nonumber\\
 % %    =& |\mathscr{S}|^{-\frac{1}{2}} \begin{bmatrix}
	% % -\frac{1}{2}\alpha_1 & \frac{1}{2} \\
	% % -\alpha_2 & 0
	% % \end{bmatrix} \varrho+\begin{bmatrix} |\mathscr{S}|^{-1/2} \vect{L}\mathcal{C}\mathcal{H}(\augstate) \\ 0 \end{bmatrix}\label{eq:varrhodot}
	% \end{align}
    where $\boldsymbol{\Psi}=\begin{bmatrix}
	-\frac{1}{2}\alpha_1 & \frac{1}{2} \\
	-\alpha_2 & 0
	\end{bmatrix} $.     
    
    After substituting $\dot{\varrho}$ in $\dot{V}_1$, one gets 
    \begin{align}
        \dot{V}_1 =& - \left(\left(|\mathscr{S}|+\epsilon\right)^{-1/2} \circ \varrho\right)^\top \mathscr{Q} \left(\left(|\mathscr{S}|+\epsilon\right)^{-1/2} \circ \varrho\right) \nonumber\\
        &+ 2 \varrho^\top \mathscr{P} \left(\left(|\mathscr{S}|+\epsilon\right)^{-1/2} \circ \vect{L}\mathcal{C}\mathcal{H}(\augstate)\right),
    \end{align}
for some $\mathscr{Q}$ satisfying $\boldsymbol{\Psi}^\top \mathscr{P} + \mathscr{P}\boldsymbol{\Psi}+\mathscr{Q}=\mathbf{0}$, where $\mathscr{Q}\succ 0$ is also constant and $\mathscr{Q} = \mathscr{Q}^\top$. 

Denote by $\lambda(\cdot)$ the eigenvalues of a matrix and recall the Cauchy-Schwarz and Rayleigh inequalities. Then, $\dot{V}_1$ satisfies
 \begin{align}
        \dot{V}_1 \leq& - \lambda_{\min}(\mathscr{Q}) \|\varrho\|^2 + 2 \lambda_{\max}(\mathscr{P}) \|\varrho\| \dfrac{\|\vect{L}\mathcal{C}\mathcal{H}(\augstate)\|}{\sqrt{\epsilon}}\nonumber\\
        \leq& - \lambda_{\min}(\mathscr{Q}) \|\varrho\|^2 + 2 \lambda_{\max}(\mathscr{P}) \|\varrho\| \dfrac{a_\mathrm{T}^{\max}}{\sqrt{\epsilon}}.\label{eq:V1dotatmax}
    \end{align}
    Whenever $\boldsymbol{\Psi}$ is Hurwitz, for every $\mathscr{Q}$, there exists a unique $\mathscr{P}$, which is the solution of the algebraic Lyapunov equation in $\mathscr{P}$ and $\mathscr{Q}$. We, then, conclude that $\dot{V}_1<0$, guaranteeing the asymptotic stability of $\mathscr{S}$ in the Lyapunov sense for an ultimate performance bound $\|\varrho\| \leq \dfrac{2 \lambda_{\max}(\mathscr{P})}{\lambda_{\min}(\mathscr{Q})} \dfrac{a_\mathrm{T}^{\max}}{\sqrt{\epsilon}}$. For $\|\varrho\|$ above this bound, $\dot{V}_1 < 0$, guaranteeing convergence to a neighborhood of the origin in finite time. Thus, the sliding mode dynamics are practically finite-time stable in the presence of an unknown (but bounded) target maneuver.

        After sliding mode is enforced, the output is governed by 
        \begin{align}\label{eq:closedloopdynamics}
            \dot{\vect{y}} = \mathcal{C}\mathcal{A}_{cl}(\augstate)\augstate\,,
        \end{align} 
        where $\mathcal{A}_{cl} = \mathcal{A}(\augstate) -\mathcal{M}(\augstate)\mathcal{P}(\augstate)$. The closed-loop matrix $\mathcal{A}_{cl}$ can be expressed as 
        \begin{align}\label{eq:closed_loop_matrix_ext}
		\mathcal{A}_{cl} = \mathcal{A}_{cl,0} + \vect{W}(\augstate) ,
		\end{align}
		where $\mathcal{A}_{cl,0} = \lim\limits_{||\vect{y}||\to \boldsymbol{0}} \mathcal{A}_{cl}(\augstate)$, and $\vect{W}:\mathbb{R}^{n}\to\mathbb{R}^{n\times n}$ is a matrix-valued function, which is continuous with respect to $\augstate$, and satisfies $\lim\limits_{||\vect{y}||\to \boldsymbol{0}}\vect{W}(\augstate)=\boldsymbol{0}.$
        
		The matrix $\mathcal{A}_{cl,0}$ can also be represented as
		% \begin{align}
		$\mathcal{A}_{cl,0} =\mathcal{A}_0 - \mathcal{B}_0{R}^{-1}\mathcal{B}^T_0\mathcal{P}_0$,
		% \end{align}
		where the matrices $\mathcal{A}_0\equiv\mathcal{A}(\vect{y}=\boldsymbol{0})$, $\mathcal{B}_0\equiv\mathcal{B}(\vect{y}=\boldsymbol{0})$ and $\mathcal{P}_0$ satisfies 
        \begin{align}
            \mathcal{A}_0^\top\mathcal{P}_0 + \mathcal{P}_0\mathcal{A}_0 - \mathcal{P}_0\mathcal{B}_0R^{-1}\mathcal{B}_0^\top\mathcal{P}_0 + \mathcal{C}^\top\vect{Q}\mathcal{C} = \boldsymbol{0}\,.
        \end{align}
        
        Now, consider the Lyapunov function $V_2(\vect{y})=\vect{y}^\top \mathcal{C}\mathcal{P}_0 \mathcal{C}^\top\vect{y}$, such that $k_1||\vect{y}||^2\leq V_2(\vect{y})\leq k_2||\vect{y}||^2,~\forall\,\vect{y}\in\mathbb{D},$
		where $\mathbb{D}$ is a bounded set in $\mathbb{R}^2$ and $k_1 \equiv \min\limits_{i\in\{1,2\},}\lambda_i\left[\mathcal{C}\mathcal{P}_0 \mathcal{C}^\top\right]$, and $k_2 \equiv \max\limits_{i\in\{1,2\}}\lambda_i \left[\mathcal{C}\mathcal{P}_0 \mathcal{C}^\top\right]$.
		% \begin{align}
		% k_1 \equiv& \min\limits_{i\in\{1,2\},}\lambda_i\left[\mathcal{C}\mathcal{P}_0 \mathcal{C}^\top\right],\\
		% k_2 \equiv& \max\limits_{i\in\{1,2\}}\lambda_i \left[\mathcal{C}\mathcal{P}_0 \mathcal{C}^\top\right].
		% \end{align}
		Substituting \eqref{eq:closed_loop_matrix_ext} in \eqref{eq:closedloopdynamics} leads to
		\begin{align}\label{eq:dynamics_mod}
		\dot{\vect{y}} = \mathcal{C}\mathcal{A}_{cl}(\augstate)\augstate = \mathcal{C}\left[\mathcal{A}_{cl,0} + \vect{W}(\augstate)\right]\augstate\,.
		\end{align}
		The total time derivative of $V_2(\vect{y})$ along the closed-loop trajectories is given by
		\begin{align}\label{eq:Vdot}
		\dot{V}_2 &=~ \dot{\vect{y}}^\top \mathcal{C}\mathcal{P}_0 \mathcal{C}^\top\vect{y} + \vect{y}^\top \mathcal{C}\mathcal{P}_0 \mathcal{C}^\top\dot{\vect{y}},\nonumber\\
		&=~ \augstate^\top\left[- \mathcal{P}_0\mathcal{B}_0R^{-1}\mathcal{B}_0^\top\mathcal{P}_0 - \mathcal{C}_0^\top\vect{Q}\mathcal{C}_0 + \boldsymbol{\Gamma}(\augstate)\right]\augstate,\nonumber\\
        &=~ \vect{y}^\top\mathcal{C}\left[- \mathcal{P}_0\mathcal{B}_0R^{-1}\mathcal{B}_0^\top\mathcal{P}_0 - \mathcal{C}_0^\top\vect{Q}\mathcal{C}_0 + \boldsymbol{\Gamma}(\augstate)\right]\mathcal{C}^\top\vect{y}\nonumber\\
        &=~ \vect{y}^\top\left[- \mathcal{C}\mathcal{P}_0\mathcal{B}_0R^{-1}\mathcal{B}_0^\top\mathcal{P}_0\mathcal{C}^\top - \vect{Q} + \mathcal{C}\boldsymbol{\Gamma}(\augstate)\mathcal{C}^\top\right]\vect{y},
		\end{align}
		with $
		\boldsymbol{\Gamma}(\augstate)=\mathcal{P}_0\vect{W}(\augstate) + \vect{W}^\top(\augstate)\mathcal{P}_0.$
		 
         Define a real-valued function $\zeta(\augstate)$ as 
		\begin{align}
		\zeta(\augstate)= \max\limits_{i\in\{1,2\}}\lambda_i\left[\mathcal{C}\boldsymbol{\Gamma}(\augstate)\mathcal{C}^\top\right],
		\end{align}
		which also satisfies
		\begin{align}\label{eq:beta_prop}
		\lim\limits_{||\vect{y}||\to \boldsymbol{0}}\zeta(\augstate)=0.
		\end{align}
		  Therefore,
		\begin{align}
		&\max\limits_{\substack{i\in\{1,2\},\\\vect{x}\in\Pi}}\lambda_i\left[- \mathcal{C}\mathcal{P}_0\mathcal{B}_0R^{-1}\mathcal{B}_0^\top\mathcal{P}_0\mathcal{C}^\top - \vect{Q} + \mathcal{C}\boldsymbol{\Gamma}(\augstate)\mathcal{C}^\top\right]\nonumber\\
		&\leq-\min\limits_{\substack{i\in\{1,2\}}}\lambda_i\left[\mathcal{C}\mathcal{P}_0\mathcal{B}_0R^{-1}\mathcal{B}_0^\top\mathcal{P}_0\mathcal{C}^\top + \vect{Q} \right]\\&\quad\quad\quad\quad\quad+\max\limits_{\substack{i\in\{1,2\},\vect{x}\in\Pi}}\lambda_i\left[ \mathcal{C}\boldsymbol{\Gamma}(\augstate)\mathcal{C}^\top\right]\nonumber\\
		&\leq-\varphi+\zeta(\augstate),
		\end{align}
		where $\varphi=\min\limits_{\substack{i\in\{1,2\}}}\lambda_i\left[\mathcal{C}\mathcal{P}_0\mathcal{B}_0R^{-1}\mathcal{B}_0^\top\mathcal{P}_0\mathcal{C}^\top + \vect{Q} \right]\,.$
        
		Following \Cref{lem:controllability} and results from infinite-horizon LQR control of LTI systems \cite{kirk2004optimal}, $\varphi>0$, and the set $\Pi\triangleq\{\vect{y}\in\mathbb{R}^2:\zeta(\augstate)<\varphi\}$ is a non-empty set due to \eqref{eq:beta_prop}. 
        
        Hence, $\forall~\vect{y}\in\Pi$,~ $\dot{V}_2<0$. More specifically, $\dot{V}_2\leq-k_3||\vect{y}||^2,$
		where, the coefficient $k_3$ is given by  
		\begin{align}\label{eq:k_3}
		k_3\equiv\min\limits_{\substack{i\in\{1,2\}}}\lambda_i\left[\mathcal{C}\mathcal{P}_0\mathcal{B}_0R^{-1}\mathcal{B}_0^\top\mathcal{P}_0\mathcal{C}^\top + \vect{Q} \right]>0.
		\end{align}
		When $\mathbb{D}=\Pi$, the origin in the output space is locally exponentially stable under the proposed guidance strategy.
    \end{proof}

\begin{figure*}[ht!]
    \centering
    \begin{subfigure}{0.3275\linewidth}
        \centering
        \includegraphics[width=\linewidth]{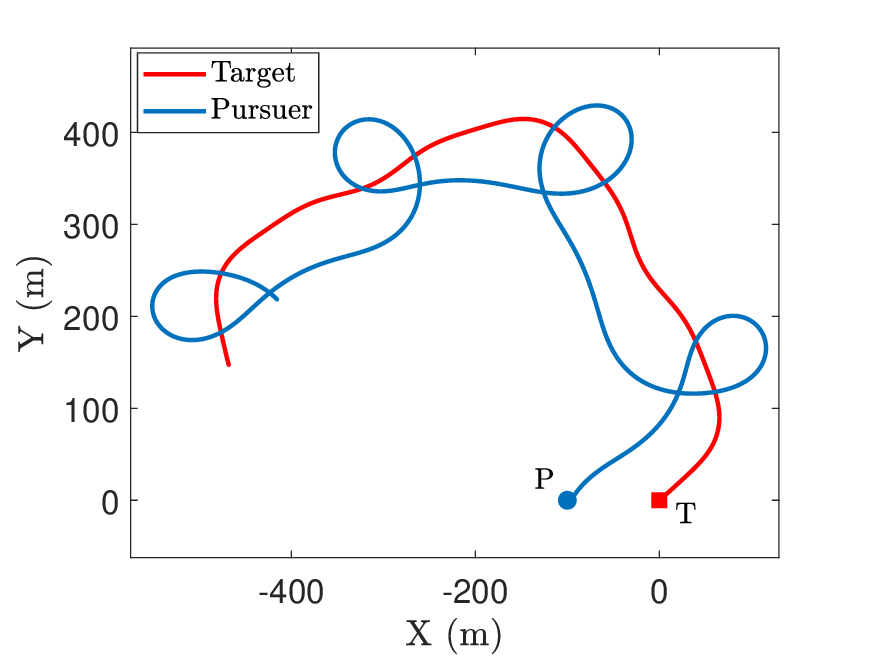}
        \caption{}
        \label{fig:MT_varrd_traj}
    \end{subfigure}
    \begin{subfigure}{0.3275\linewidth}
        \centering
        \includegraphics[width=\linewidth]{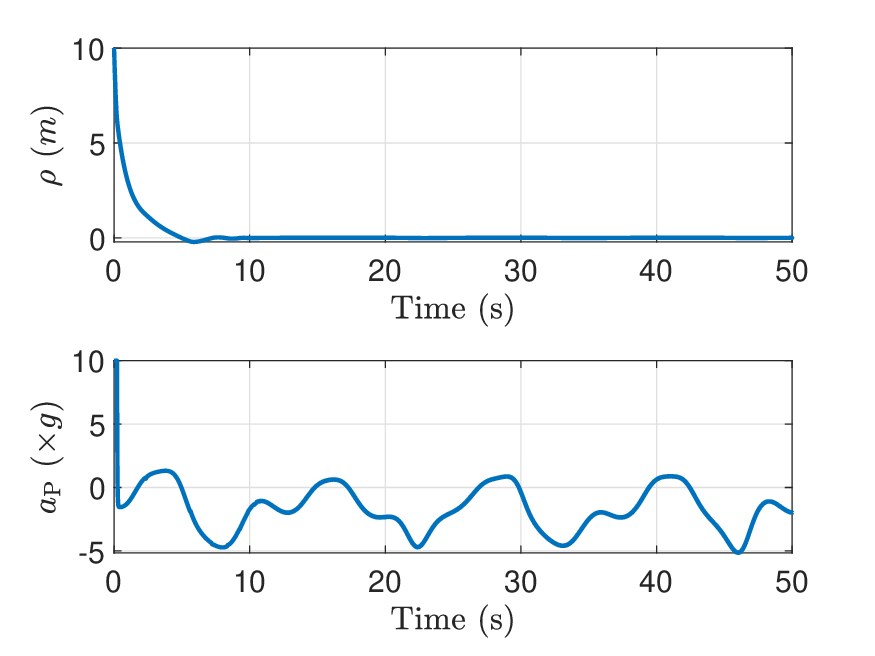}
        \caption{}
        \label{fig:MT_varrd_rhoAp}
    \end{subfigure}
    \begin{subfigure}{0.3275\linewidth}
        \centering
        \includegraphics[width=\linewidth]{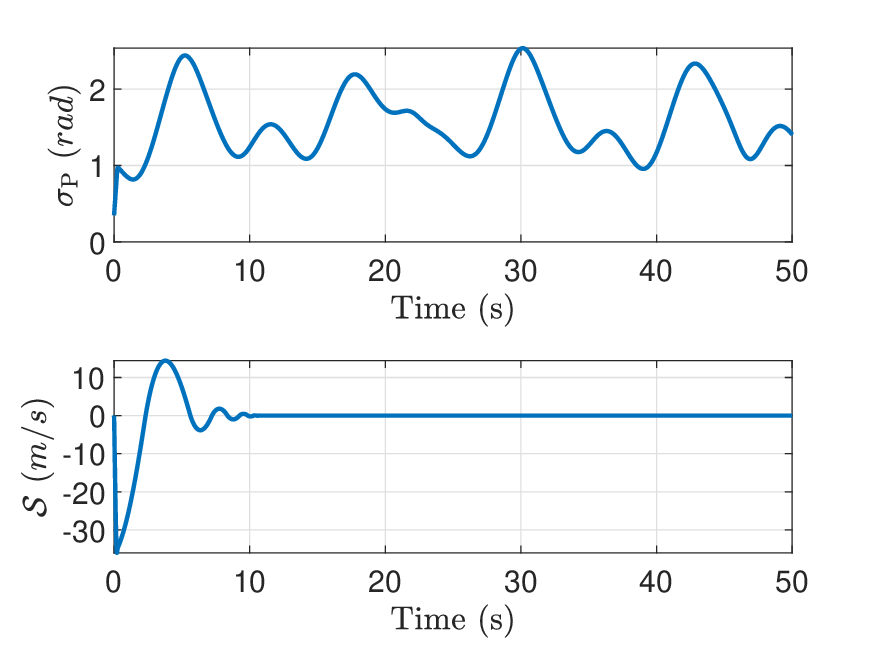}
        \caption{}
        \label{fig:MT_varrd_sigPS}
    \end{subfigure}
    \caption{Simulation results for Case 1.}
    \label{fig:illustrative_simresultsc1}
\end{figure*}
\begin{figure*}[ht!]
    % ----------------
    \begin{subfigure}{0.3275\linewidth}
        \centering
        \includegraphics[width=\linewidth]{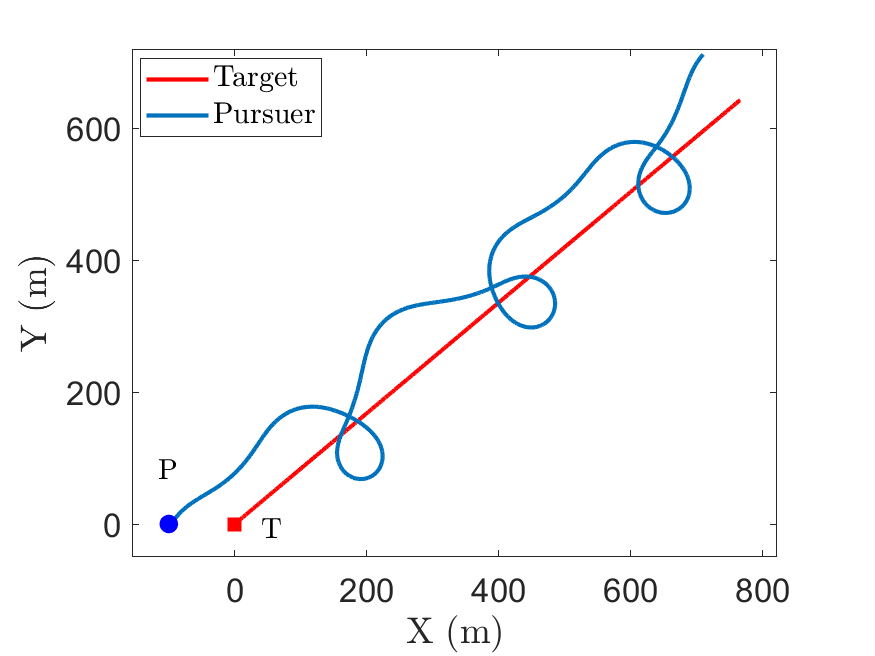}
        \caption{}
        \label{fig:CVT_varrd_traj}
    \end{subfigure}
    \begin{subfigure}{0.3275\linewidth}
        \centering
        \includegraphics[width=\linewidth]{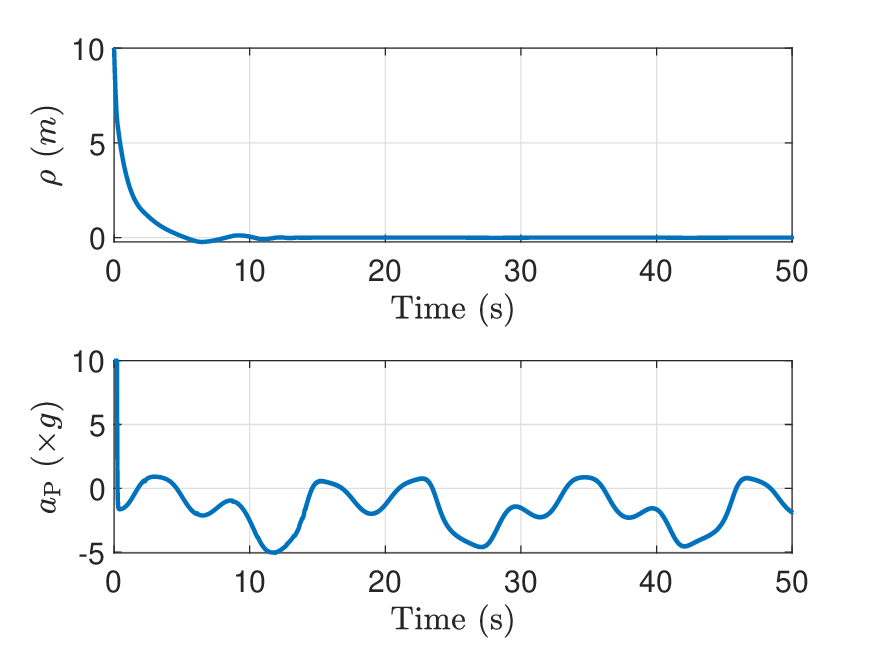}
        \caption{}
        \label{fig:CVT_varrd_rhoAp}
    \end{subfigure}
    \begin{subfigure}{0.3275\linewidth}
        \centering
        \includegraphics[width=\linewidth]{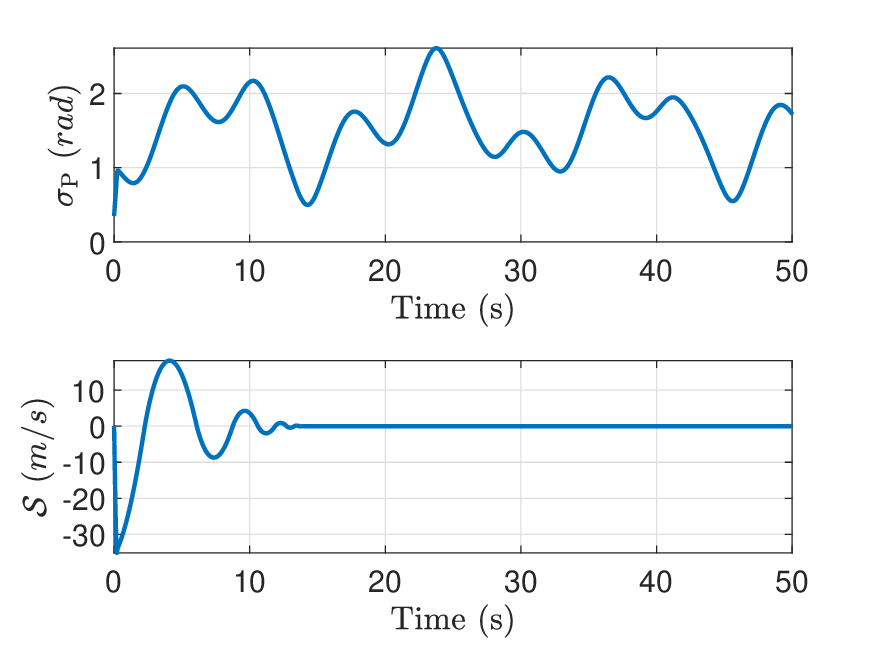}
        \caption{}
        \label{fig:CVT_varrd_sigPS}
    \end{subfigure}
    \caption{Simulation results for Case 2.}
    \label{fig:illustrative_simresultsc2}
\end{figure*}
\begin{figure*}[ht!]
    % ----------------
    \begin{subfigure}{0.3275\linewidth}
        \centering
        \includegraphics[width=\linewidth]{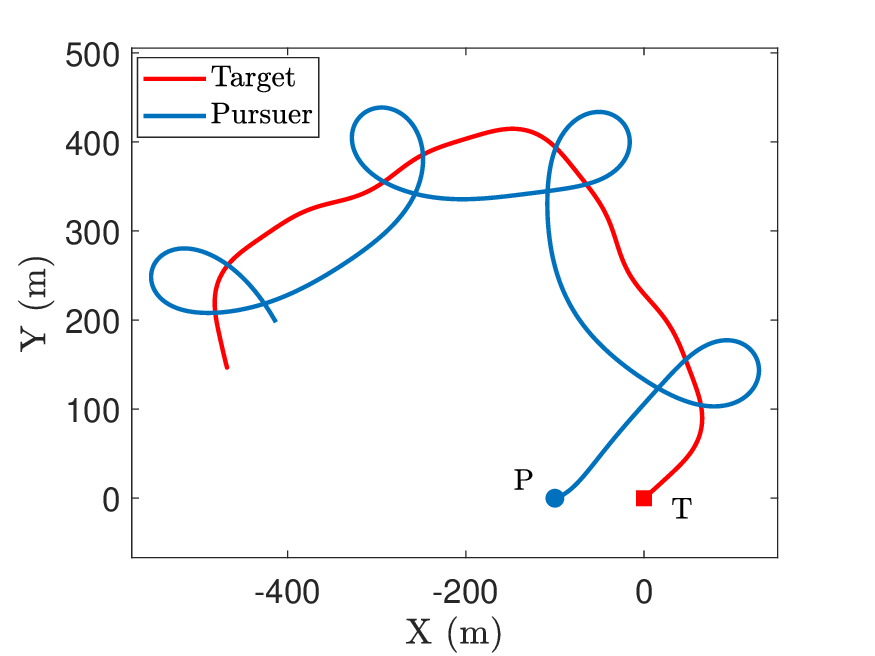}
        \caption{}
        \label{fig:MT_constrd_traj}
    \end{subfigure}
    \begin{subfigure}{0.3275\linewidth}
        \centering
        \includegraphics[width=\linewidth]{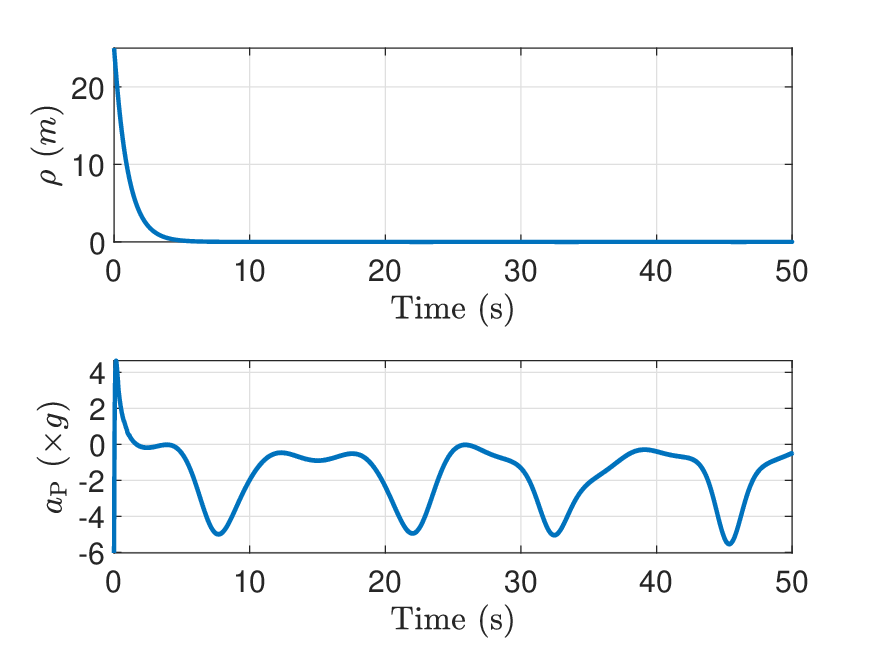}
        \caption{}
        \label{fig:MT_constrd_rhoAp}
    \end{subfigure}
    \begin{subfigure}{0.3275\linewidth}
        \centering
        \includegraphics[width=\linewidth]{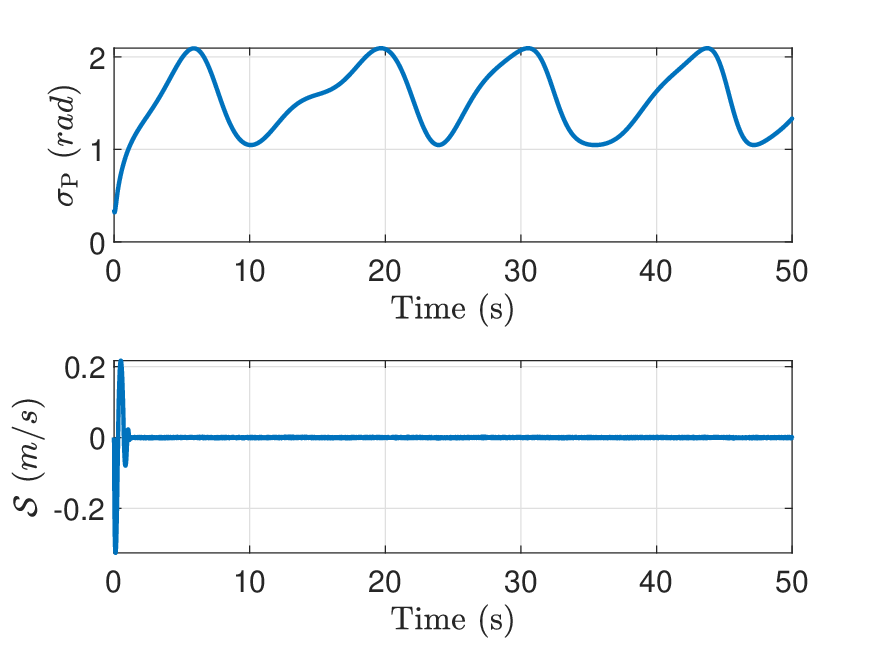}
        \caption{}
        \label{fig:MT_constrd_sigPS}
    \end{subfigure}
    \caption{Simulation results for Case 3.}
    \label{fig:illustrative_simresultsc3}
\end{figure*}
\begin{figure*}[ht!]
    % ----------------
    \begin{subfigure}{0.3275\linewidth}
        \centering
        \includegraphics[width=\linewidth]{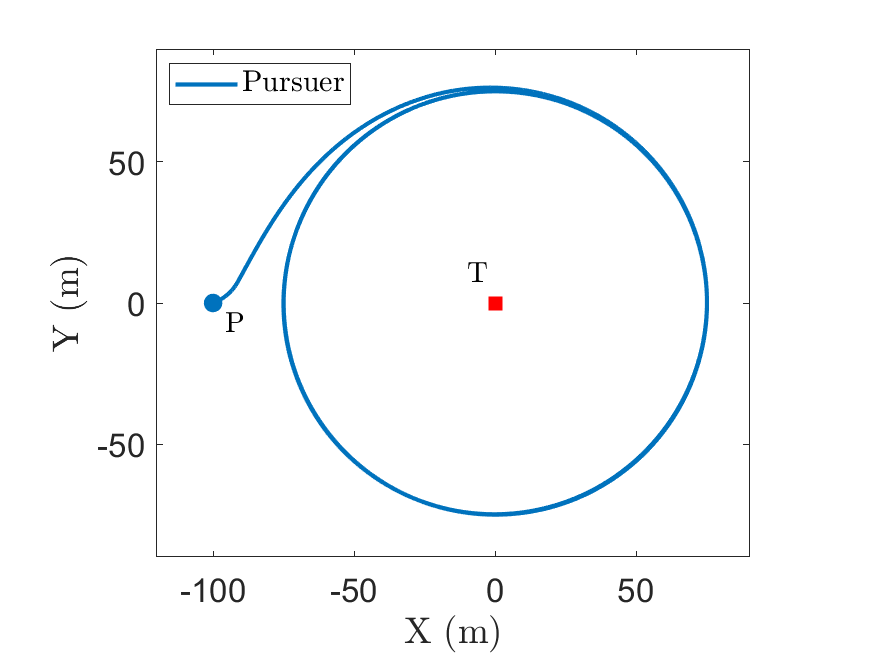}
        \caption{}
        \label{fig:ST_constrd_traj}
    \end{subfigure}
    \begin{subfigure}{0.3275\linewidth}
        \centering
        \includegraphics[width=\linewidth]{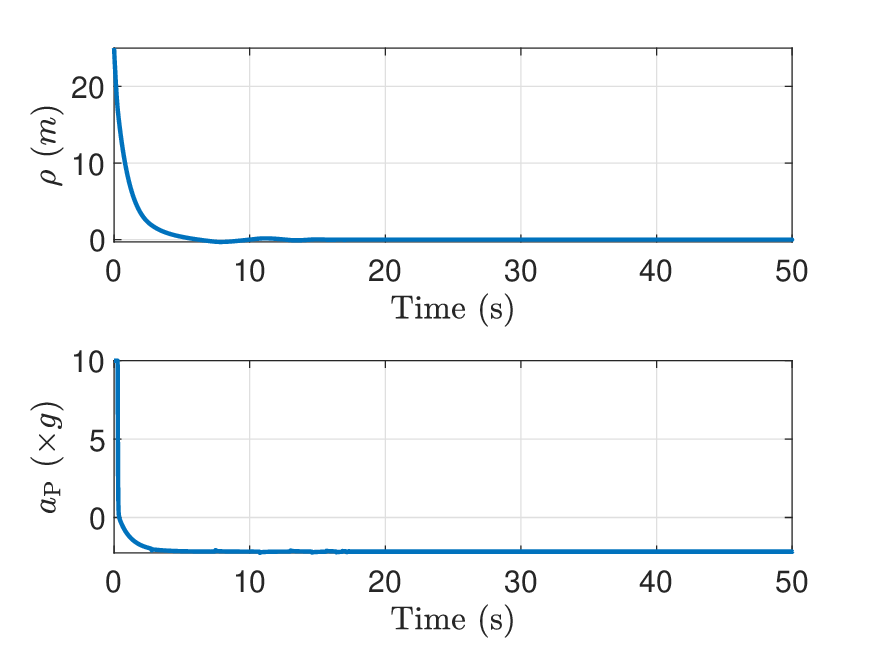}
        \caption{}
        \label{fig:ST_constrd_rhoAp}
    \end{subfigure}
    \begin{subfigure}{0.3275\linewidth}
        \centering
        \includegraphics[width=\linewidth]{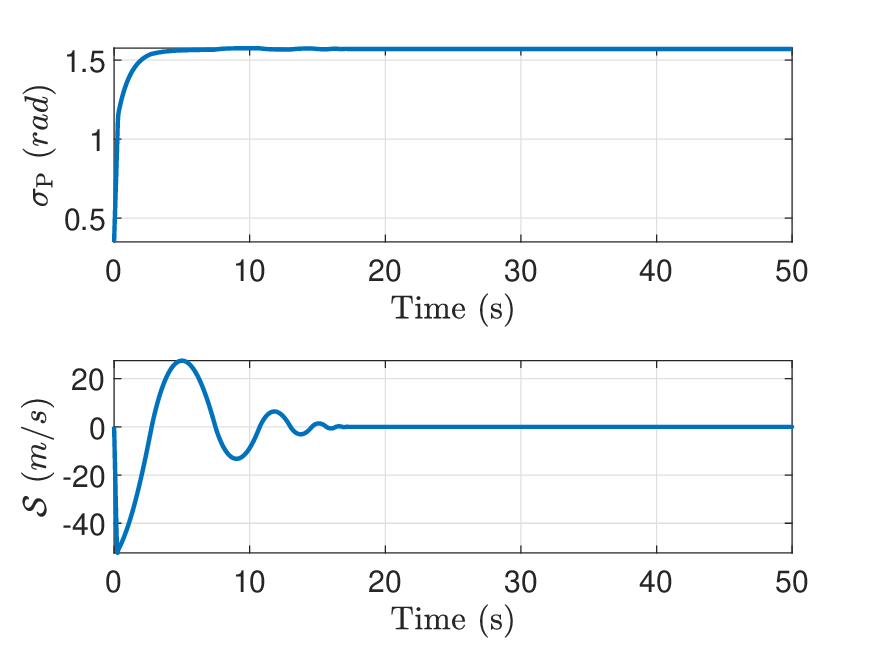}
        \caption{}
        \label{fig:ST_constrd_sigPS}
    \end{subfigure}
    \caption{Simulation results for Case 4.}
    \label{fig:illustrative_simresultsc4}
\end{figure*}

\begin{remark}
        If $\mathbf{g}(\mathbf{x})$ is contained in $\vect{h}(\vect{x})$, then $\|\vect{L}\mathcal{C}\mathcal{H}(\augstate)\|\leq a_\mathrm{T}^{\max}+r_d^{\max}\left(v_\mathrm{P}+v_\mathrm{T}\right)^2+\ddot{r}_d^{\max}$ since $\dot{\theta}$ is also upper bounded by $v_\mathrm{P}+v_\mathrm{T}$. Therefore, \eqref{eq:V1dotatmax} should be replaced by
        \begin{align}
        \dot{V}_1 \leq& - \lambda_{\min}(\mathscr{Q}) \|\varrho\|^2 \nonumber\\&+ 2 \lambda_{\max}(\mathscr{P}) \|\varrho\| \dfrac{a_\mathrm{T}^{\max}+r_d^{\max}\left(v_\mathrm{P}+v_\mathrm{T}\right)^2+\ddot{r}_d^{\max}}{\sqrt{\epsilon}}
    \end{align}
    without affecting the stability proof, which reiterates the essence of \Cref{rmk:gxhx}. However, the ultimate performance bound may change due to extra unknown information. 
    \end{remark}

\section{Simulations}
In this section, we evaluate the performance of the proposed robust near-optimal nonlinear target enclosing guidance strategy for various target types, such as stationary, constant-velocity and maneuvering targets for a time-varying enclosing geometry. %Additionally, we will also be comparing its performance to existing enclosing guidance strategies like Sinha et. al. in \cite{Sinha2022}.

% \subsection{Illustrative Study}
We consider a pursuer moving at 40 m/s while the target may be either mobile or stationary. The pursuer's maximum lateral acceleration is limited to $\pm10g$ due to actuator constraints, with $g$ being the acceleration due to gravity. The simulation results consist of trajectory plots, lateral acceleration commands, the range error profile, $\rho(t)$, the pursuer lead angle $\sigma_\mathrm{P}(t)$, and the sliding manifold $\mathscr{S}$. The pursuer is initially located 100 m away from the target, which is assumed to start from the origin at the beginning of the engagement. The initial LOS angle for the pursuer-target engagement is chosen to 0$^\circ$. For scenarios with a moving target, the target’s initial heading angle is chosen to be 40$^\circ$, whereas the pursuer starts with a heading angle of 20$^\circ$. The design parameters to implement the proposed target enclosing guidance strategy are $\mathbf{Q} = \text{\rm diag}(10^8,~10^8)$, $R = 4\times 10^{4}$, $\alpha_1 = 10$, $\alpha_2 = 10$, and $\beta = 0.5$.

\Crefrange{fig:illustrative_simresultsc1}{fig:illustrative_simresultsc4} contains the illustrative simulation results for four different scenarios for the desired enclosing profile and target maneuver as mentioned in \Cref{tab:initcond}.
% \begin{table}[!h]
%     \centering
%     \begin{tabular}{c|c|c}
%         \hline
%         & \textbf{$r_d(t)~(m)$} & \textbf{$a_T(t)~(m/s^2)$}\\\hline
%         \textbf{Case 1} & $75 + 2 \sin(t) + 15\cos(t)$ & $1.5 - 5\cos(0.2\pi t)\sin(0.1\pi t)$ \\
%         \textbf{Case 2} & $75 + 2 \sin(t) + 15\cos(t)$ & $0$\\
%         \textbf{Case 3} & $75$ & $1.5 - 5\cos(0.2\pi t)\sin(0.1\pi t)$\\
%         \textbf{Case 4} & $75$ & $0$\\\hline
%     \end{tabular}
%     \caption{Parameters for results in Fig. \ref{fig:illustrative_simresultsc1}-\ref{fig:illustrative_simresultsc4}.}
%     \label{tab:initcond}
% \end{table}
\begin{table}[!h]
    \centering
    \begin{tabular}{@{} lcc @{}}
        \toprule
        & \textbf{$r_d(t)\,(m)$} & \textbf{$a_\mathrm{T}(t)\,(m/s^2)$} \\ 
        \midrule
        \textbf{Case 1} & $75 + 2 \sin(t) + 15\cos(t)$ & $1.5 - 5\cos(0.2\pi t)\sin(0.1\pi t)$ \\
        \textbf{Case 2} & $75 + 2 \sin(t) + 15\cos(t)$ & $0$ \\
        \textbf{Case 3} & $75$ & $1.5 - 5\cos(0.2\pi t)\sin(0.1\pi t)$ \\
        \textbf{Case 4} & $75$ & $0$ \\
        \bottomrule
    \end{tabular}
    \caption{Parameters for simulations.}
    \label{tab:initcond}
\end{table}

Cases 1 and 2 consider a time-varying desired relative distance $r_d(t)$ with respect to a maneuvering and constant velocity target, respectively. Furthermore, Cases 3 and 4 focus on fixed-distance stand-off tracking of a maneuvering and a stationary target, respectively. As $r_d(t)$ is a fixed constant in Case 4, the consequent scenario reduces to encirclement of a stationary target. 

It can be noted from \Crefrange{fig:illustrative_simresultsc1}{fig:illustrative_simresultsc4} that the proposed guidance strategy is successful in enclosing a target regardless of its  maneuvering level for each case in \Cref{tab:initcond}. Additionally, one can also note from \Cref{fig:MT_varrd_rhoAp,fig:CVT_varrd_rhoAp,fig:MT_constrd_rhoAp,fig:ST_constrd_rhoAp} that the relative range error profile $\rho(t)$ converges to near zero within finite time and the commanded lateral acceleration $a_{\rm P}$ remains well-defined for each case. Similar behavior can be noted for the sliding manifold profile $\mathscr{S}(t)$ from \Cref{fig:MT_varrd_sigPS,fig:CVT_varrd_sigPS,fig:MT_constrd_sigPS,fig:ST_constrd_sigPS}, showing convergence after a transient of small duration.

As Case 4 represents a target encirclement scenario, the proposed lateral acceleration demand converges to a constant value upon the convergence of $\rho(t)\to0$ as seen from  \Cref{fig:ST_constrd_rhoAp}. Furthermore, it can be observed from \Cref{fig:ST_constrd_sigPS} that the pursuer's lead angle $\sigma_{\rm P}$ also converges to $\pi/2$ as $\rho(t)\to0$. %Note that from these representative cases, we see that  $\sigma_\mathrm{P}\in\{0,\pm\pi\}$ 

% We also test the performance of the proposed generalized enclosing guidance strategy with existing enclosing guidance scheme presented in \cite{Sinha2022}. For the purpose of the comparison, we consider a target moving at a speed of $20 m/s$ with a maneuver profile of $a_T = 2 - 5\sin(0.1\pi t)$. Furthermore, we also consider the 

% \begin{figure*}
%     \centering
%     \begin{subfigure}{0.3275\linewidth}
%         \centering
%         \includegraphics[width=\linewidth]{Figures/comp_MT_varRd_traj.eps}
%         \caption{}
%         \label{fig:comp_MT_varrd_traj}
%     \end{subfigure}
%     \begin{subfigure}{0.3275\linewidth}
%         \centering
%         \includegraphics[width=\linewidth]{Figures/comp_MT_varRd_rho_aP.eps}
%         \caption{}
%         \label{fig:comp_MT_varrd_rhoAp}
%     \end{subfigure}
%     \begin{subfigure}{0.3275\linewidth}
%         \centering
%         \includegraphics[width=\linewidth]{Figures/comp_MT_varRd_sigP_S.eps}
%         \caption{}
%         \label{fig:comp_MT_varrd_sigPS}
%     \end{subfigure}
%     \caption{Comparison results with Sinha et. al. \cite{Sinha2022}.}
%     \label{fig:comparisonstudy}
% \end{figure*}
% \subsection{Comparison Study}

\section{Conclusions}
In this work, we proposed a robust near-optimal nonlinear guidance framework for enclosing a general maneuvering target within arbitrary smooth closed geometries, which is a generalization of classic circumnavigation methods. The proposed design requires only relative state measurements and avoids reliance on explicit curvature or target maneuver information, thereby enhancing its applicability under uncertain or adversarial scenarios. By leveraging the SDRE framework, the proposed approach allows synthesizing locally optimal feedback commands for nonlinear pursuer-target engagement. To further strengthen robustness, the proposed design was augmented with an integral sliding mode manifold, ensuring resilience against disturbances that may otherwise deviate the system from its nominal trajectory.  Future work may involve multi-agent (both pursuers and targets) scenarios. %\tr{Simulation results confirm the effectiveness of the proposed method, highlighting its capability to achieve target enclosing with improved robustness and reduced control effort relative to existing state-of-the-art approaches.}

\bibliographystyle{ieeetr}
\bibliography{references}

\end{document}